\documentclass[12pt]{article}
\usepackage{graphicx}
\usepackage[figuresright]{rotating}
\newcommand{\beq}{\begin{equation}}
\newcommand{\eeq}{\end{equation}}
\newcommand{\be}{\begin{eqnarray}}
\newcommand{\ee}{\end{eqnarray}}
\begin{document}
\rightline{RUB-TPII-12/99}
\rightline{hep-ph/9909541}
\vspace{.3cm}
\begin{center}
\begin{large}
{\bf Flavor asymmetry of polarized antiquark distributions
and semi-inclusive DIS} \\[1cm]
\end{large}
\vspace{1.4cm}
{\bf B.\ Dressler}$^{\rm a, 1}$, {\bf K.\ Goeke}$^{\rm a, 2}$, 
{\bf M.V.\ Polyakov}$^{\rm a, b, 3}$, {\bf and C. Weiss}$^{\rm a, 4}$  
\\[1.cm]
$^{a}${\em Institut f\"ur Theoretische Physik II,
Ruhr--Universit\"at Bochum, \\ D--44780 Bochum, Germany} 
\\
$^{b}${\em Petersburg Nuclear Physics Institute, Gatchina, \\
St.Petersburg 188350, Russia} 
\end{center}
\vspace{1.5cm}
\begin{abstract}
\noindent
The $1/N_c$--expansion of QCD suggests large flavor asymmetries of the 
polarized antiquark distributions in the nucleon. This is confirmed 
by model calculations in the large--$N_c$ limit (chiral quark--soliton 
model), which give sizable results for $\Delta\bar u(x) - \Delta\bar d (x)$
and $\Delta\bar u(x) + \Delta\bar d (x) - 2 \Delta \bar s (x)$. 
We compute the contributions of these flavor asymmetries 
to the spin asymmetries in hadron production in semi-inclusive 
deep--inelastic scattering. We show that the large flavor asymmetries 
predicted by the chiral quark--soliton model are consistent with the 
recent HERMES data for spin asymmetries in charged hadron production.
\end{abstract}
\vfill
\rule{5cm}{.2mm} \\
\noindent
{\footnotesize $^{\rm 1}$ E-mail: birgitd@tp2.ruhr-uni-bochum.de} \\
{\footnotesize $^{\rm 2}$ E-mail: goeke@tp2.ruhr-uni-bochum.de} \\
{\footnotesize $^{\rm 3}$ E-mail: maximp@tp2.ruhr-uni-bochum.de} \\
{\footnotesize $^{\rm 4}$ E-mail: weiss@tp2.ruhr-uni-bochum.de}
\newpage
\section{Introduction}
The study of polarized parton distributions in the nucleon
presents a major challenge to both experiment and theory. Particularly 
subtle issues are the polarized antiquark distributions, and
the precise flavor decomposition of the polarized quark and antiquark
distributions, including the strangeness contributions. Knowledge of 
the latter is a prerequisite {\it e.g.}\ for the identification of 
the gluon contribution to the proton spin \cite{COMPASS}.
\par
Traditional inclusive lepton scattering experiments, {\it i.e.}, 
measurements of the polarized structure functions, do not allow 
to directly distinguish 
between the polarized quark-- and antiquark distributions. Rather, 
quark-- and antiquark contributions have to be identified from the study 
of scaling violations, which implies a considerable loss of 
accuracy \cite{GRSV96}. 
Moreover, the flavor decomposition can only be studied by way of 
comparing experiments
with different targets, typically proton and light nuclei (deuteron, 
helium), which is rendered difficult by nuclear binding effects.
Much more direct access to the individual quark-- and
antiquark distributions is possible in polarized semi-inclusive 
DIS (deep--inelastic scattering),
where one measures {\it e.g.}\ the spin asymmetry of the cross 
section for producing a certain hadron in the fragmentation of 
the struck quark or antiquark in the target.  Such measurements have 
recently been performed
by the SMC \cite{SMC98} and HERMES \cite{HERMES99} experiments.
The unpolarized quark and antiquark fragmentation functions needed 
in the QCD description of these asymmetries can be
measured independently in $e^+ e^-$--annihilation into 
hadrons \cite{BKK97}, and 
also in hadron production in unpolarized DIS
off the nucleon \cite{EMC89,Geiger98}.
\par
There have also been attempts to estimate the flavor asymmetry of the
polarized antiquark distributions theoretically, using models for the 
structure of the nucleon. A large flavor asymmetry of the polarized 
antiquark distribution was first obtained in a calculation
of the quark-- and antiquark distributions at a low scale in 
the large--$N_c$ limit ($N_c$ is the number of colors), 
where the nucleon can be described as a soliton of an effective 
chiral theory \cite{DPPPW96,DPPPW97,Dressler98}. 
The unpolarized quark-- and antiquark 
distributions \cite{DPPPW96,DPPPW97,PPGWW98}, as well as the polarized 
distributions of quarks plus antiquarks \cite{DPPPW96,DPPPW97,WK99}
calculated in this approach are in good agreement with the 
standard parametrizations obtained from fits to inclusive DIS 
data \cite{GRSV96}. In the $1/N_c$--expansion 
the isovector polarized distributions are leading compared to the isoscalar
ones, and calculations in the chiral quark--soliton model, using standard 
parameters, give an isovector antiquark distribution, 
$\Delta \bar u(x) - \Delta \bar d (x)$, considerably larger 
than the isoscalar one, $\Delta \bar u(x) + \Delta \bar d (x)$. 
Such a large polarized antiquark flavor asymmetry 
should lead to observable effects in semi-inclusive spin asymmetries 
as measured {\it e.g.}\ by the HERMES experiment. 
It should be noted that the same approach describes
well the observed violation of the Gottfried sum rule 
\cite{NMC94,Gottfried67,Kumano97} and the recent
data for the $x$--dependence of the flavor asymmetry of the 
unpolarized antiquark distribution from Drell--Yan pair 
production \cite{E866} and semi-inclusive DIS \cite{HERMES98},
see Refs.\cite{PPGWW98,Dressler98,WW99} for details.
\par
Recently, the polarized antiquark flavor asymmetry has been estimated
in approaches which generalize the meson cloud picture of DIS
off the nucleon to the polarized 
case \cite{Fries98,BoreskovKaidalov98}. It is known that pion exchange 
contributions to DIS off the 
nucleon provide a qualitative explanation for the observed flavor asymmetry 
of the unpolarized antiquark distribution \cite{Sullivan72,cloud}. In 
Ref.\cite{Fries98} the authors considered the contribution of 
polarized rho meson exchange to the polarized antiquark distributions
in the nucleon and obtained an estimate of the flavor asymmetry considerably
smaller than the large--$N_c$ result of Refs.\cite{DPPPW96,DPPPW97}.
\par
In this paper we offer new arguments in favor of a large
flavor asymmetry of the polarized antiquark distributions.
Our main points are two: First, on the theoretical side, we comment
on the estimates of the polarized flavor asymmetry in the
meson exchange picture in Refs.\cite{Fries98,BoreskovKaidalov98}. 
Specifically, we argue that the polarized rho meson exchange contributions 
considered in Ref.\cite{Fries98} are not the dominant contributions
within that approach, so that a small value obtained for this contribution
does not imply smallness of the total polarized antiquark flavor 
asymmetry. Second, we study the implications of the flavor asymmetry 
of the polarized antiquark distributions for the spin asymmetries 
measured in hadron production in semi-inclusive DIS.
Combining information on the polarized quark and antiquark
distributions available from inclusive DIS with the large--$N_c$ model
calculation of the polarized flavor asymmetries, we make quantitative 
predictions for the spin asymmetries in semi-inclusive pion, kaon, and 
charged particle production. We discuss the sensitivity of these 
observables to the flavor asymmetries of the polarized antiquark 
distributions. With the quantitative
estimate of $\Delta \bar u (x) - \Delta \bar d(x)$ from the chiral 
quark soliton model we obtain a sizable contribution of the flavor 
asymmetry to semi-inclusive spin asymmetries.
Actually, incorporating the effects of the large flavor asymmetry
our results fit well the recent HERMES data for spin asymmetries
in semi-inclusive charged hadron production \cite{HERMES99}. We 
discuss the assumptions made in the analysis of the HERMES data 
in Ref.\cite{HERMES99}, and argue that they are too restrictive
and might have led to a bias in favor of a small flavor asymmetry. 
We also make predictions for experiments with the possibility to 
measure spin asymmetries of individual charged hadrons
($\pi^+, \pi^-, K^+, K^-$), which could be feasible at 
HERMES or CEBAF.
\par
\section{Flavor asymmetry of the polarized antiquark
distribution in the large--$N_c$ limit}
{\it Quark-- and antiquark distributions in the large--$N_c$ limit.}
A very useful tool for connecting QCD with the hadronic 
world is the theoretical limit of a large number of colors.
Qualitatively speaking, at large $N_c$ QCD becomes equivalent to a theory 
of mesons, with baryons appearing as solitonic excitations \cite{Witten}.
The $1/N_c$--expansion allows to classify baryon and meson masses, 
weak and strong characteristics in a model--independent way; usually
the estimates agree surprisingly well with phenomenology. 
One example are the 
isovector and isoscalar axial coupling constants of the nucleon,
which are of the order
\be
g_A^{(3)} &\sim& N_c ,
\hspace{2cm} 
g_A^{(0)} \;\; \sim \;\; N_c^0 ,
\label{g_A_N_c}
\ee
in qualitative agreement with the numerical values extracted from
experiments, $g_A^{(3)} \approx 1.25$ and $g_A^{(0)} \approx 0.3$.
The same technique has been applied to the parton distributions in the 
nucleon at a low normalization point \cite{DPPPW96}. There one finds
that the isoscalar unpolarized and the isovector polarized distributions
of quarks and antiquarks are leading in the $1/N_c$--expansion, 
while the respective
other flavor combinations, the isovector unpolarized and isoscalar 
polarized distributions,
appear only in the next--to--leading order. More precisely, 
at large $N_c$ the distributions scale as
\be 
\mbox{leading:} \;\;\;\;\;\;\;
\left. 
\begin{array}{l} 
u(x) + d(x), \;\; \bar u(x) + \bar d(x) 
\\[.5cm] 
\Delta u(x) - \Delta d(x), \;\; \Delta \bar u(x) - \Delta \bar d(x)
\end{array}
\right\}
&\sim&
N_c^2 \,\, f(N_c x),
\label{Nc_large}
\\[.5cm]
\mbox{subleading:} \;\;\;\;\;\;\;
\left. 
\begin{array}{l} 
u(x) - d(x), \;\; \bar u(x) - \bar d(x) 
\\[.5cm] 
\Delta u(x) + \Delta d(x), \;\; \Delta \bar u(x) + \Delta \bar d(x)
\end{array}
\right\}
&\sim&
N_c \,\, f(N_c x) ,
\label{Nc_small}
\ee
where $f(y)$ is a stable function in the large $N_c$--limit, which
depends on the particular distribution considered. Note that the
large $N_c$--behavior of the polarized quark-- and antiquark 
distributions is related to that of the corresponding axial coupling 
constants, Eq.(\ref{g_A_N_c}) by the sum rules for the first moments 
of these distributions (Bjorken and Ellis--Jaffe sum rules).
\par
It is interesting that the isovector polarized antiquark distribution 
is parametrically larger than the isoscalar one. While the 
$1/N_c$--expansion is only a parametric estimate, it is nevertheless
an indication that $\Delta \bar u (x) - \Delta \bar d (x)$ could be 
also numerically large. This is indeed confirmed by model calculations
(see below). 
\par
{\it Polarized vs.\ unpolarized antiquark flavor asymmetry.}
We would like to briefly comment on the assumptions
about the polarized antiquark flavor asymmetry made in
the recent analysis of the HERMES data for semi-inclusive 
DIS \cite{HERMES99}. From the point of view of 
the $1/N_c$--expansion the flavor asymmetry of the
polarized antiquark distribution is parametrically larger than 
that of the unpolarized ones. Thus, the assumption 
in the fit of Ref.\cite{HERMES99} of proportional flavor asymmetry 
in the polarized and unpolarized antiquark distributions, namely 
$\Delta \bar u (x) / \bar u (x) = \Delta \bar d (x) / \bar d (x)$,
is inconsistent with the $1/N_c$--expansion and appears unnatural. 
The consequences of this assumption can be seen more clearly
if one notes that it implies
\be
\Delta \bar u(x) - \Delta \bar d(x)
&=&  \left[ \bar u(x) - \bar d(x) \right]
\frac{\Delta \bar u(x) + \Delta \bar d(x)}{\bar u(x) + \bar d(x)} .
\ee
The ratio of the isoscalar polarized to the isoscalar
unpolarized distribution on the R.H.S.\ is always less
than unity, which follows from the probabilistic interpretation
of the leading--order distributions considered here. Thus, with
the above assumption the polarized antiquark flavor
asymmetry can never be larger numerically than the unpolarized
one. Consequently, a fit under the assumption 
$\Delta \bar u (x) / \bar u (x) = \Delta \bar d (x) / \bar d (x)$
cannot be regarded as a real alternative to the reference fit 
assuming zero polarized antiquark flavor asymmetry, 
$\Delta \bar u (x) - \Delta \bar d (x) = 0$.
In this sense it seems that the analysis of Ref.\cite{HERMES99}
contained an implicit bias in favor of a small polarized antiquark 
flavor asymmetry.
\par
{\it Model calculation of quark and antiquark distributions
at a low normalization point.}
In order to make quantitative estimates of the parton distributions at a 
low normalization point one needs to supplement the large--$N_c$ limit 
with some dynamical information. It is known that at low energies the 
behavior of strong interactions is largely determined by the spontaneous 
breaking of chiral symmetry. A concise way to summarize the implications 
of this non-perturbative phenomenon is by way of an effective field theory, 
valid at low energies. Such a theory has been derived ``microscopically''
within the framework of the instanton description of the QCD vacuum, 
which provides a dynamical explanation for the breaking of chiral
symmetry in QCD \cite{DP86}. It can be expressed in terms of an effective
Lagrangian describing quarks with a dynamical mass, interacting with
pions, which appear as Goldstone bosons in the spontaneous breaking of 
chiral symmetry (here $x$ denotes the space--time coordinates):
\beq
L_{\rm eff} \;\; = \;\; \bar \psi(x) \left[ i \gamma^\mu \partial_\mu
- M \, U^{\gamma_5} (x) \right] \psi(x) ,
\label{L_eff} 
\eeq
\beq
U^{\gamma_5} (x) \;\; \equiv \;\; \frac{1 + \gamma_5}{2} U(x) + 
\frac{1 - \gamma_5}{2} U^\dagger (x) .
\eeq
Here $U(x)$ is a unitary matrix containing the Goldstone boson
degrees of freedom, which can be parametrized as
\be
U(x) &=& \frac{1}{F_\pi} \left[ \sigma (x) + i \tau^a \pi^a (x) \right],
\hspace{2cm} \sigma^2 + (\pi^a )^2  \;\; = \;\; 
F_\pi^2 .
\ee
($F_\pi = 93 \, {\rm MeV}$ is the weak pion decay constant).
The effective theory is valid up to an ultraviolet cutoff, whose
value is of the order $600\,{\rm MeV}$ \cite{DP86}. 
\par
In the large--$N_c$ limit the nucleon in the effective theory defined by 
Eq.(\ref{L_eff}) is described by a classical pion field
which binds the quarks (chiral quark--soliton model) \cite{DPP88}. 
The field is of ``hedgehog'' form; in the nucleon rest frame it is 
given by
\be
U_{\rm cl} ({\bf x}) &=& 
\exp\left[ i \frac{x^a \tau^a}{r} P(r)\right],
\hspace{1.5cm} r \; \equiv \; |{\bf x}|, 
\label{hedge}
\ee
where $P(r)$ is called the profile function; $P(0) = -\pi$, and
$P(r) \rightarrow 0$ for 
$r \rightarrow \infty$. Nucleon states with definite spin/isospin
and momentum emerge after quantizing the collective rotations and
translations of the soliton. The parton distributions in the nucleon
at large $N_c$ can be computed by summing over the contributions of quark 
single--particle states in the background field; the normalization point
is of the order of the ultraviolet cutoff of the effective theory,
$\mu \approx 600 \, {\rm MeV}$ (see Refs.\cite{DPPPW96,DPPPW97} 
for details).
\par
{\it Gradient expansion of the isovector polarized antiquark 
distribution.}
Analytic expressions for the parton distributions in the large--$N_c$
nucleon can be obtained in the theoretical limit of large soliton 
size, where one can perform an expansion of the sum over quark levels 
in gradients of the classical pion field, Eq.(\ref{hedge}) \cite{DPPPW96}.
The isovector polarized antiquark distribution in leading--order gradient 
expansion is given by
\be 
\Delta \bar u(x) - \Delta \bar d(x)
&=&
\frac{F_\pi^2 M_N}{3} \int_{-\infty}^\infty \frac{d\xi}{2\pi}
\frac{\cos M_N \xi x}{\xi} \;
\int d^3 y \; {\rm tr}\,\left[ \tau^3 (-i) 
U_{\rm cl} ({\bf y} + \xi {\bf e}_3)
U^\dagger_{\rm cl} ({\bf y}) \right] 
\nonumber \\
&=& 
\frac{4 M_N}{3} \int_{-\infty}^\infty \frac{d\xi}{2\pi}
\frac{\cos M_N \xi x}{\xi} \;
\int d^3 y \; \pi^3_{\rm cl} ({\bf y} + \xi {\bf e}_3) 
\sigma_{\rm cl} ({\bf y}) ,
\label{gradient}
\ee
where $M_N$ denotes the nucleon mass, ${\bf e}_3$ the three--dimensional 
unit vector in the 3--direction, and $\tau^3$ the isospin Pauli 
matrix.\footnote{We remark that, when combined with the corresponding 
gradient expansion expression for the isovector polarized quark distribution, 
the first moment of Eq.(\ref{gradient}) reproduces the well--known 
expression for the gradient expansion of the isovector axial coupling 
constant, $g_A^{(3)}$, of the large--$N_c$ nucleon.}
The reason why we are interested in this theoretical limit
is that it allows to make explicit the dependence of the polarized
antiquark distribution on the classical chiral fields of the soliton. 
This will be useful for the discussion in Section \ref{sec_cloud}.
Aside from this, comparison with the result of exact numerical 
calculations shows that the leading--order gradient expansion, 
Eq.(\ref{gradient}), gives already a very realistic numerical estimate 
of the isovector polarized antiquark distribution \cite{DPPPW97}. 
\begin{figure}[t]
\begin{center}
\includegraphics[width=8cm,height=8cm]{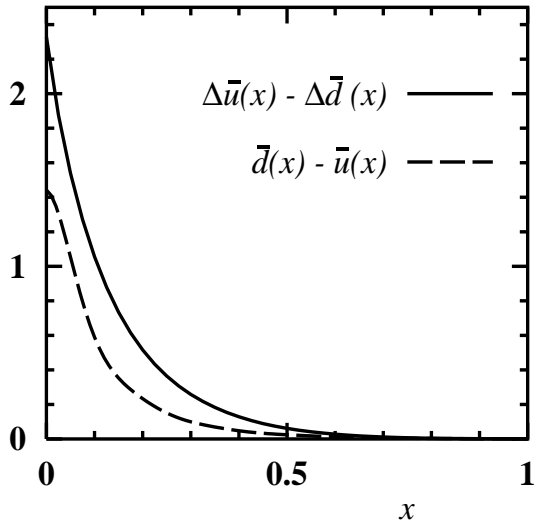}
\end{center}
\caption[]{\it \underline{Solid line:} The isovector polarized antiquark 
distribution at the
low normalization point ($\mu = 600\, {\rm MeV}$), as obtained
from the chiral soliton model of the nucleon (for details see 
the text). \underline{Dashed line:} The isovector unpolarized antiquark 
distribution calculated in Ref.\cite{PPGWW98}.}
\label{fig_delta}
\end{figure}
\par
{\it Numerical result for $\Delta \bar u(x) - \Delta\bar d (x)$.}
In the numerical estimates of semi-inclusive asymmetries
in Section \ref{sec_semi} we shall use not the gradient 
expansion formula, Eq.(\ref{gradient}), but a more accurate 
numerical estimate obtained by adding the bound--state level 
contribution and using interpolation--type formula to estimate 
the continuum contribution \cite{DPPPW96}. This result for the 
distribution is shown in Fig.\ref{fig_delta}. One sees that the flavor
asymmetry of the polarized antiquark distribution is numerically
larger than the unpolarized one \cite{PPGWW98}, in agreement with 
the fact that it is leading in the $1/N_c$--expansion. Note 
that the unpolarized antiquark flavor asymmetry calculated in 
this approach is in agreement with the results of the analysis 
of the E866 Drell--Yan data \cite{E866} as well as with the
HERMES measurements in semi-inclusive DIS \cite{HERMES98} (for details, 
see Refs.\cite{PPGWW98,Dressler98,WW99}).
\par
{\it Including strangeness.} For a realistic description of semi-inclusive
spin asymmetries one has to take into account the polarized 
distribution of strange quarks and antiquarks in the nucleon. Within 
the large--$N_c$ description of the nucleon it is possible to include
strangeness by extending the effective low--energy theory
in the chiral limit, Eq.(\ref{L_eff}) to three quark flavors and treating 
corrections due to the finite strange current quark mass perturbatively.
In this approach the nucleon is described by embedding the $SU(2)$ hedgehog, 
Eq.(\ref{hedge}), in the $SU(3)$ flavor space, and quantizing
its flavor rotations in the full $SU(3)$ flavor space. Flavor symmetry 
breaking can then be included perturbatively by computing matrix elements 
of symmetry--breaking operators between $SU(3)$--symmetric nucleon 
states \cite{Blotz93}.
For our estimates here we limit ourselves to the simplest case of unbroken 
$SU(3)$ symmetry ($m_s = 0$). In this case collective quantization
of the $SU(3)$ rotations of the soliton leads to a simple relation
between the flavor--octet and triplet polarized antiquark 
distributions, namely
\be
\Delta \bar u (x) + \Delta \bar d (x) - 2 \Delta \bar s (x)
&=&
\frac{3 F - D}{F + D} 
\left[ \Delta \bar u (x) - \Delta \bar d (x) \right] .
\label{8_from_3}
\ee
The value of $F/D$ has been estimated in the chiral quark soliton 
model \cite{antidek}.
When one regards the radius of the soliton as a free parameter
(in reality it is determined from minimizing the energy of the soliton), 
the result for $F/D$ interpolates between the $SU(6)$ quark model 
value (for small soliton size), $F/D = 2/3$, and the value obtained in the
Skyrme model (for large soliton size), $F/D = 5/9$.\footnote{It is
interesting to note that in this approach the value for $F/D$ is 
one-to-one related to the isoscalar axial coupling constant, $g_A^{(0)}$,
see Ref.\cite{antidek}.} Using the value $F/D = 5/9$ corresponding to the 
limit of large soliton size, the ratio in Eq.(\ref{8_from_3}) 
comes to $3/7$. We shall use the relation Eq.(\ref{8_from_3}) with this
value in our estimates of semi-inclusive spin 
asymmetries in Section \ref{sec_semi}.
\section{On the flavor asymmetry of the polarized antiquark 
distribution in the meson cloud picture}
\label{sec_cloud}
A widely used phenomenological model for the flavor asymmetry of the 
unpolarized antiquark distributions in the nucleon is the meson cloud 
picture \cite{cloud}. Recently there have been attempts to estimate
also the polarized antiquark flavor asymmetry in this approach
\cite{Fries98,BoreskovKaidalov98}. In particular, the authors
of Ref.\cite{Fries98} obtained an estimate for 
$\Delta \bar u (x) - \Delta \bar d (x)$ more than an order of magnitude 
smaller than the result of the large--$N_c$ model calculation, 
Fig.\ref{fig_delta}. This striking disagreement may lead to the 
impression that the present theoretical understanding
of the polarized antiquark flavor asymmetry is very poor.
In this situation we consider it helpful to briefly comment on the
estimates within the meson cloud picture. Specifically, we want
to show how the very small estimate obtained in Ref.\cite{Fries98}
could be reconciled with our large--$N_c$ result. 
To avoid misunderstandings, we stress at this point that, in spite of
many superficial similarities, the meson cloud picture described here
differs in many crucial respects from the large--$N_c$ approach
(more on this below), so the two approaches should not be confused.
\par
The meson cloud picture of DIS off the
nucleon assumes that the nucleon can be described as a ``bare'' nucleon,
characterized by flavor--symmetric quark-- and antiquark distributions, 
and a ``cloud'' of virtual mesons. The flavor asymmetry of the antiquark 
distributions
is then attributed to processes in which the hard probe couples to such a 
virtual meson. For instance, the sign of the observed unpolarized antiquark 
asymmetry in the proton,  $\bar d(x) - \bar u(x) > 0$, can qualitatively be 
explained by the photon coupling to a pion in the ``cloud''
(Sullivan mechanism \cite{Sullivan72}), if one takes into
account that the emission of a $\pi^+$ by the proton, with transition 
to a neutron or $\Delta^0$ intermediate state, is favored
compared to that of a $\pi^-$, which is possible only by a transition 
to a $\Delta^{++}$ state, as illustrated in
Fig.\ref{fig_clpi}.\footnote{For a discussion of the role of
the $\pi N N$ and $\pi N \Delta$ form factors in quantitative 
estimates of these contributions, see Ref.\cite{KFS96}.}
%
%
\refstepcounter{figure}
\label{fig_clpi}
\begin{center}
\includegraphics[width=10cm,height=4cm]{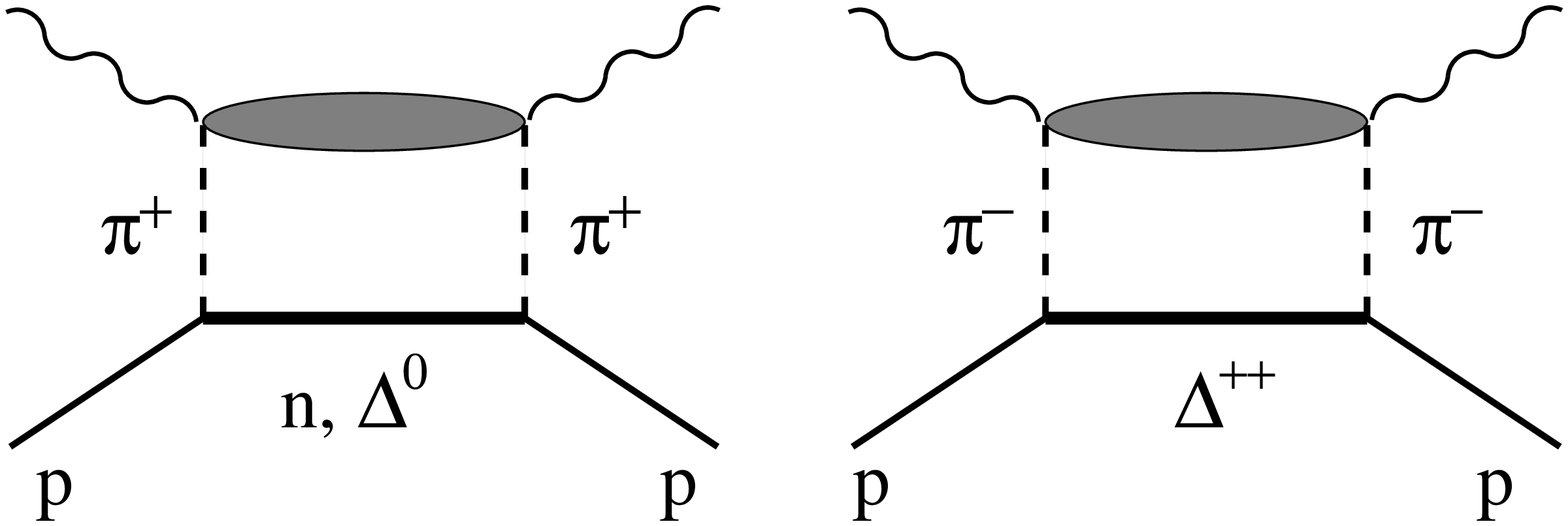}
\end{center}
Figure~\thefigure : {\it The Sullivan mechanism contributing to the
unpolarized antiquark flavor asymmetry in the proton, 
$\bar u(x) - \bar d (x)$.}
\vspace{.5cm}
\par
The simple Sullivan mechanism involving the pion ``cloud'' does not 
contribute to the polarized asymmetry, which has often been taken as an 
argument in favor of the smallness of this asymmetry.
Recently, Boreskov and Kaidalov \cite{BoreskovKaidalov98} made the interesting 
observation that at small $x$ a sizable polarized antiquark 
asymmetry is generated by the interference of the amplitudes 
for the photon coupling to a pion and to a rho meson emitted by the
nucleon, as shown schematically in Fig.\ref{fig_clpirh}. \\
%
%
\refstepcounter{figure}
\label{fig_clpirh}
\begin{center}
\includegraphics[width=10cm,height=4cm]{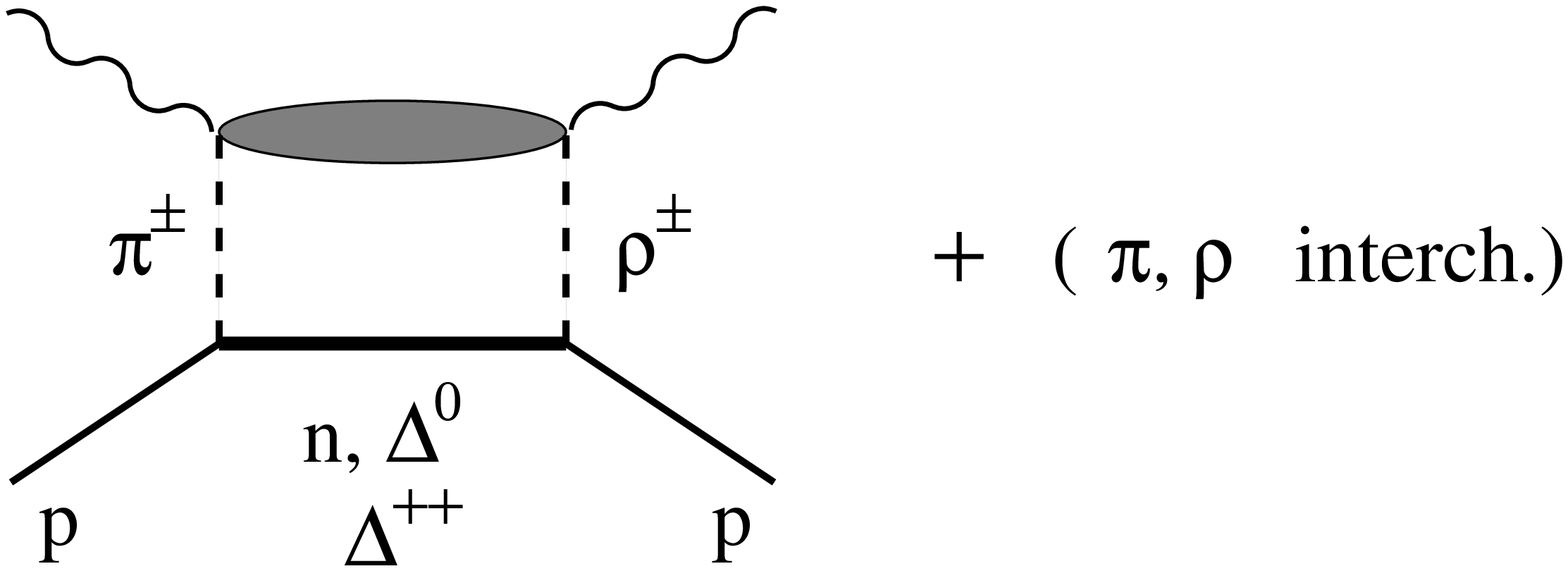}
\end{center}
Figure~\thefigure : {\it The pi--rho interference contributions
to the isovector polarized structure function at small $x$
considered in Ref.\cite{BoreskovKaidalov98}.}
\\[.5cm]
This type of exchange 
corresponds to the leading Regge cut contributing to the imaginary 
part of the high--energy photon--nucleon scattering amplitude.
Previously, Fries and Sch\"afer \cite{Fries98}
had considered the Sullivan--type
contribution from polarized rho meson exchange to the polarized antiquark 
asymmetry, $\Delta\bar u (x) - \Delta\bar d (x)$, at larger values of $x$
({\it i.e.}, not restricted to small $x$), see Fig.\ref{fig_clrhrh}. \\
%
%
\refstepcounter{figure}
\label{fig_clrhrh}
\begin{center}
\includegraphics[width=5cm,height=4cm]{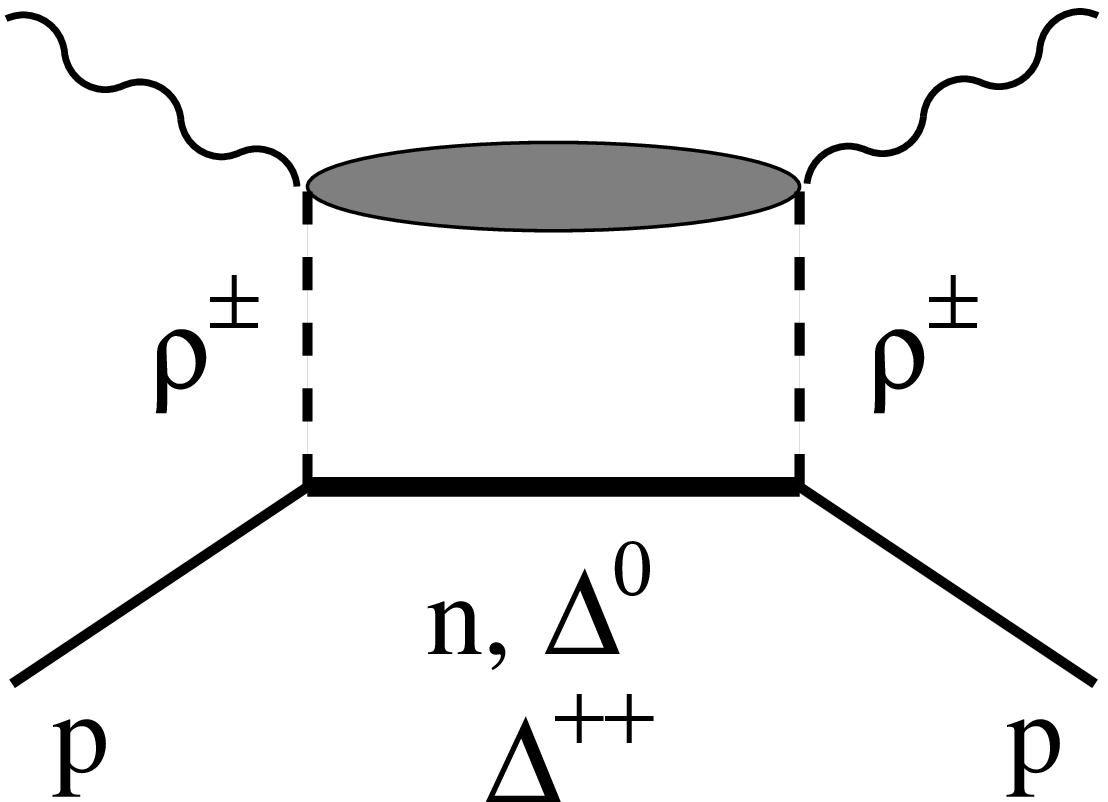}
\end{center}
Figure~\thefigure : {\it The polarized rho meson exchange 
contributions to $\Delta \bar u(x) - \Delta \bar d (x)$ 
considered in Ref.\cite{Fries98}.}
\\[.5cm]
They obtained a strikingly small contribution to 
$\Delta\bar u (x) - \Delta\bar d (x)$,
roughly two orders of magnitude smaller than the result of the calculation 
in the large--$N_c$ limit shown
in Fig.\ref{fig_delta}. It should be noted, however, that within 
the meson cloud picture contributions of type of Fig.\ref{fig_clrhrh}
are not special. In fact, one can obtain a contribution to the
polarized antiquark distribution already from the exchange of
spin--$0$ mesons. To see this it is instructive to take a look at the 
gradient expansion of the isovector polarized antiquark distribution in 
large--$N_c$ limit, Eq.(\ref{gradient}). Although the fields in 
Eq.(\ref{gradient}) are the classical chiral fields of the soliton,
and no direct interpretation of this expression in terms of simple 
meson exchange diagrams is possible, it can provide some qualitative 
insights as to which quantum numbers can contribute 
to the polarized flavor asymmetry. Eq.(\ref{gradient}) 
suggests that, in the language of the meson cloud model, 
a contribution to the polarized antiquark asymmetry at average values 
of $x$ should come already from the interference of pion and ``sigma meson'' 
exchange, as illustrated in Fig.\ref{fig_clpisi}.
\\
\refstepcounter{figure}
\label{fig_clpisi}
\begin{center}
\includegraphics[width=10cm,height=4cm]{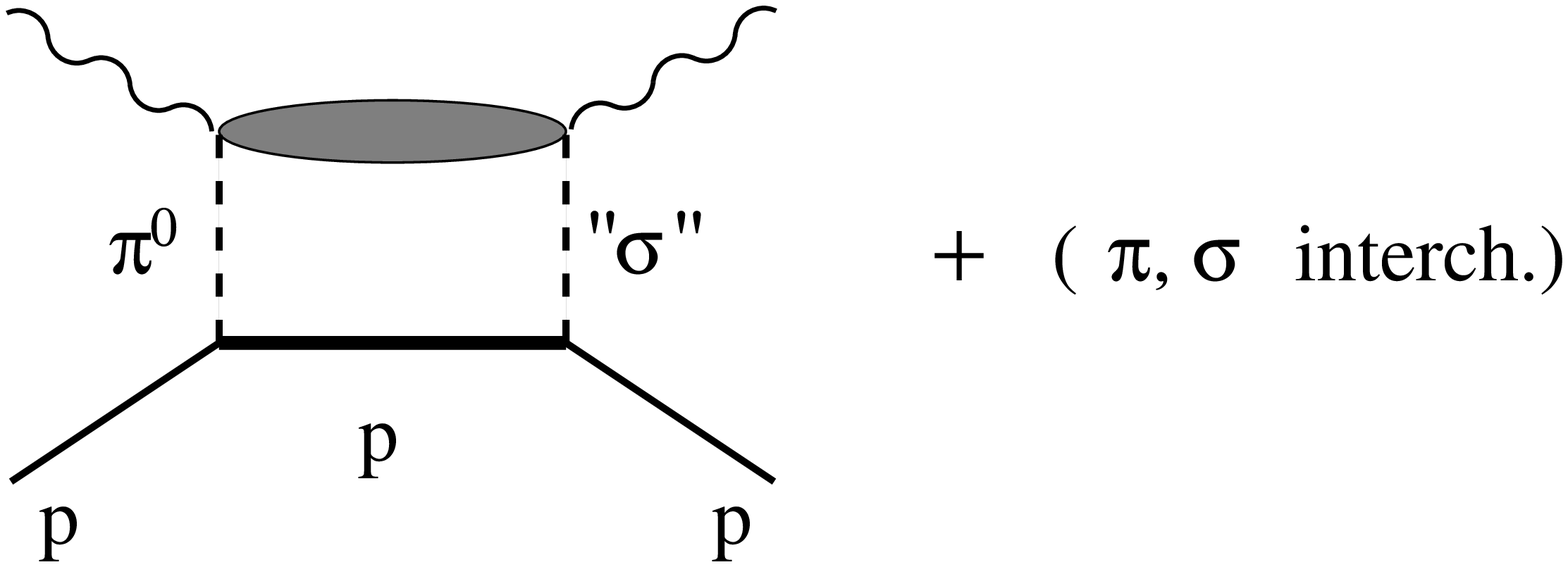}
\end{center}
Figure~\thefigure : {\it Schematic illustration of the ``pi--sigma'' 
interference likely to give a large contribution to
$\Delta \bar u(x) - \Delta \bar d (x)$ in the meson cloud
picture (for details, see the text).}
\\[.5cm]
Given the large mass of the rho meson compared to the pion, 
and the additional suppression due to need to have a polarized
rho meson, it is not difficult to imagine that this interference
contribution could give a much larger contribution to
$\Delta\bar u (x) - \Delta\bar d (x)$ than the Sullivan--type 
polarized rho meson exchange of Fig.\ref{fig_clrhrh}.
\par
It is difficult in QCD to meaningfully speak about exchanges of mesons 
other than pions, which play a special role as Goldstone bosons of 
spontaneously broken chiral symmetry, and as mediators of strong 
interactions at long distances.\footnote{It should be noted that
even in the case of pure pion exchange, which can in principle
be properly defined in the soft--pion limit, the notion of 
meson exchange contributions to the nucleon structure functions
presents severe conceptual difficulties, since in graphs 
of the type of Fig.\ref{fig_clpi} the typical momenta of the 
exchanged pions are not small, but can run up 
to momenta of the order of $\sim 1 \, {\rm GeV}$ \cite{KFS96}.}
In a ``pure'' pion cloud picture, contributions of the type indicated 
in Fig.\ref{fig_clpisi} should be referred to as interference between 
the photon scattering off a flavor--symmetric ``bare'' nucleon and a pion 
in the ``cloud''.
\par
We emphasize that the large--$N_c$ results for the flavor asymmetries
of the antiquark distributions in the nucleon cannot generally
be interpreted as single meson exchange diagrams such as 
Figs.\ref{fig_clpi}--\ref{fig_clpisi}. This can be seen from the fact 
that the large--$N_c$ limit of individual meson exchange 
diagrams typically gives rise to a wrong large--$N_c$ behavior of the 
resulting quark and antiquark distributions, different from 
Eqs.(\ref{Nc_large}) and (\ref{Nc_small}). The large--$N_c$ approach 
avoids the arbitrary separation of nucleon structure 
functions into ``core'' and ``cloud'' contributions. At the same
time, however, this approach retains the possibility of describing genuine 
Goldstone boson exchange contributions at large distances. For 
example, the large--$N_c$ result for the helicity skewed quark distribution 
correctly reproduces the singularity found in this distribution in the 
chiral limit in QCD, which can be attributed to Goldstone boson 
exchange \cite{Penttinen98}.
\section{Semi-inclusive spin asymmetries}
\label{sec_semi}
In leading--order QCD the spin asymmetry of the cross section
for semi-inclusive production of a hadron of type $h$ in the 
deep--inelastic scattering of a virtual photon off a hadronic 
target is given by
\be
A_1^h (x, z; Q^2) &=& \frac{\sum_a e_a^2 \; \Delta q_a (x, Q^2) 
\; D_a^h (z, Q^2)}
{\sum_b e_b^2 \; q_b (x, Q^2 ) \; D_b^h (z, Q^2)} 
\left[ \frac{1 + R(x, Q^2 )}{1 + \gamma^2 } \right] ,
\label{A1_x_z}
\ee
where $x$ is the Bjorken variable, $Q^2 = -q^2$ the photon virtuality, 
and $q_a (x, Q^2)$ and $\Delta q_a (x, Q^2)$, 
denote, respectively, the unpolarized and polarized quark 
and antiquark distributions in the target, at the scale $Q^2$. 
The sum over $a, b$ implies the sum over light quark flavors as 
well as over quarks/antiquarks:
\[
a, b \;\; = \;\; \left\{ u, \bar u , d, \bar d , s , \bar s \right\} .
\]
Furthermore, $D_a (z, Q^2)$ denotes the quark and antiquark 
fragmentation functions, describing the probability for the 
struck quark of type $a$ to fragment into a hadron of type $h$ 
with fraction $z$ of its longitudinal momentum.
Finally, in Eq.(\ref{A1_x_z}) $R (x, Q^2 ) = \sigma_L / \sigma_T$
is the usual ratio of the total longitudinal to the transverse photon 
cross section, and $\gamma = 2 x M_N / \sqrt{Q^2}$ is a kinematical factor.
\par
Instead of the spin asymmetry for fixed $z$, Eq.(\ref{A1_x_z}), one 
usually considers the so-called integrated asymmetry, which is defined 
as
\be
A_1^h (x; Q^2 ) &=& \frac{\sum_a e_a^2 \; \Delta q_a (x; Q^2) \; 
D_a^h (Q^2 )}
{\sum_b e_b^2 \; q_b (x; Q^2 ) \; D_b^h (Q^2)}
\left[ \frac{1 + R(x, Q^2 )}{1 + \gamma^2 } \right] ,
\label{A1_x}
\ee
where
\be
D_a^h (Q^2) &=& \int_{z_{\rm min}}^1 dz\, D_a^h (z; Q^2 ) .
\label{D}
\ee
Here $z_{\rm min} > 0$ is a cutoff which ensures that the 
observed hadron was in fact produced by fragmentation of the
struck quark in the target (suppression of target 
fragmentation) \cite{Geiger98}.
\par
Due to the presence of the (anti--) quark fragmentation functions
in the expression for the asymmetries Eqs.(\ref{A1_x_z}) and (\ref{A1_x})
the quark and antiquark distributions in the target enter with 
different coefficients.
This is different from the inclusive spin asymmetry, which at the same 
level of approximation is given by
\be
A_1 (x, Q^2) &=& \frac{\sum_a e_a^2 \; \Delta q_a (x, Q^2)}
{\sum_b e_b^2 \; q_b (x, Q^2 )} 
\left[\frac{1 + R(x, Q^2 )}{1 + \gamma^2 } \right] .
\label{A1_incl}
\ee
[Up to kinematical factors this quantity is equal to the ratio
of polarized to unpolarized structure functions, 
$g_1 (x, Q^2 ) / F_1 (x, Q^2 )$]. In the following it will be convenient 
to rewrite the expression for the semi-inclusive asymmetry, Eq.(\ref{A1_x}),
in such a way as to explicitly separate the contributions of those
combinations of parton distributions which are known well from 
inclusive DIS, from others which discriminate
between quark and antiquark distributions. We write
\be
A_1^h (x; Q^2 ) &=& \left[
A_{1, \, u}^h \; + \; A_{1, \, d}^h \; + \; A_{1, \, s}^h
\; + \; A_{1, \, 0}^h \; + \; A_{1, \, 3}^h \; + \; A_{1, \, 8}^h 
\right] (x; Q^2 ) ,
\label{A_decomposition}
\ee
where the contributions are defined as 
(we omit the $Q^2$--dependence for brevity)
\be
A_{1, \, u}^h (x) &=& X_u^h 
\left[ \Delta u (x) + \Delta\bar u (x) \right]
\hspace{1cm} \mbox{(analogously for $d, s$)} ,
\\[1ex]
A_{1, \, 0}^h (x) &=& X_0^h \left[
\Delta \bar u(x) + \Delta \bar d(x) + \Delta \bar s(x) \right] ,
\\[1ex]
A_{1, \, 3}^h (x) 
&=& X_3^h \left[\Delta \bar u(x) - \Delta \bar d(x) \right] ,
\\[1ex]
A_{1, \, 8}^h (x) &=& X_8^h \left[\Delta \bar u(x) + \Delta \bar d(x) 
- 2\Delta \bar s(x) \right] .
\ee
The coefficients $X_u^h, \ldots X_8^h$ are given by the following
combination of quark charges and quark fragmentation functions:
\be
X_u^h (x) &=& \frac{e_u^2 D_u^h}{Y}
\hspace{1cm} \mbox{(analogously for $d, s$)} ,
\label{X_u}
\\[1ex]
X_0^h &=& \frac{1}{3 Y} \left[ -e_u^2 (D^h_u - D^h_{\bar u})
 - e_d^2 (D^h_d - D^h_{\bar d}) - e_s^2 (D^h_s - D^h_{\bar s}) \right] ,
\label{X_0}
\\[1ex]
X_3^h &=& \frac{1}{2 Y} \left[ -e_u^2 (D^h_u - D^h_{\bar u})
 + e_d^2 (D^h_d - D^h_{\bar d}) \right] ,
\label{X_3}
\\[1ex]
X_8^h &=& \frac{1}{6 Y} \left[ -e_u^2 (D^h_u - D^h_{\bar u})
 - e_d^2 (D^h_d - D^h_{\bar d}) + 2 e_s^2 (D^h_s - D^h_{\bar s}) \right] ,
\label{X_8}
\ee
with
\be
Y &=& \frac{1 + \gamma^2 }{1 + R(x, Q^2 )} 
\; \sum_a e_a^2 \; q_a (x; Q^2 ) \; D_a^h (Q^2) .
\ee
The terms $A_{1, \, u}^h, A_{1, \, d}^h$ and $A_{1, \, s}^h$ contain the 
contributions of the 
sum of quark and antiquark distributions, which appear also
in the inclusive polarized spin asymmetry (polarized structure functions), 
Eq.(\ref{A1_incl}), and can therefore be measured independently
in DIS. [Actually, in DIS with proton or nuclear targets 
one is able to measure directly only two flavor combinations of these 
three distributions; the third one can be inferred using $SU(3)$ 
symmetry arguments.] The term 
$A_{1, \, 0}^h$ in Eq.(\ref{A_decomposition}) contains the 
flavor--singlet polarized antiquark distribution.
The terms $A_{1, \, 3}^h$ and $A_{1, \, 8}^h$, finally,
are proportional to the flavor--nonsinglet (triplet and octet, respectively)
combinations of the polarized antiquark distributions, which do not
contribute to inclusive DIS and are therefore
left essentially unconstrained in the parametrizations of 
parton distributions derived from fits to inclusive data \cite{GRSV96}.
\begin{figure}[t]
\begin{center}
\includegraphics[width=8cm,height=8cm]{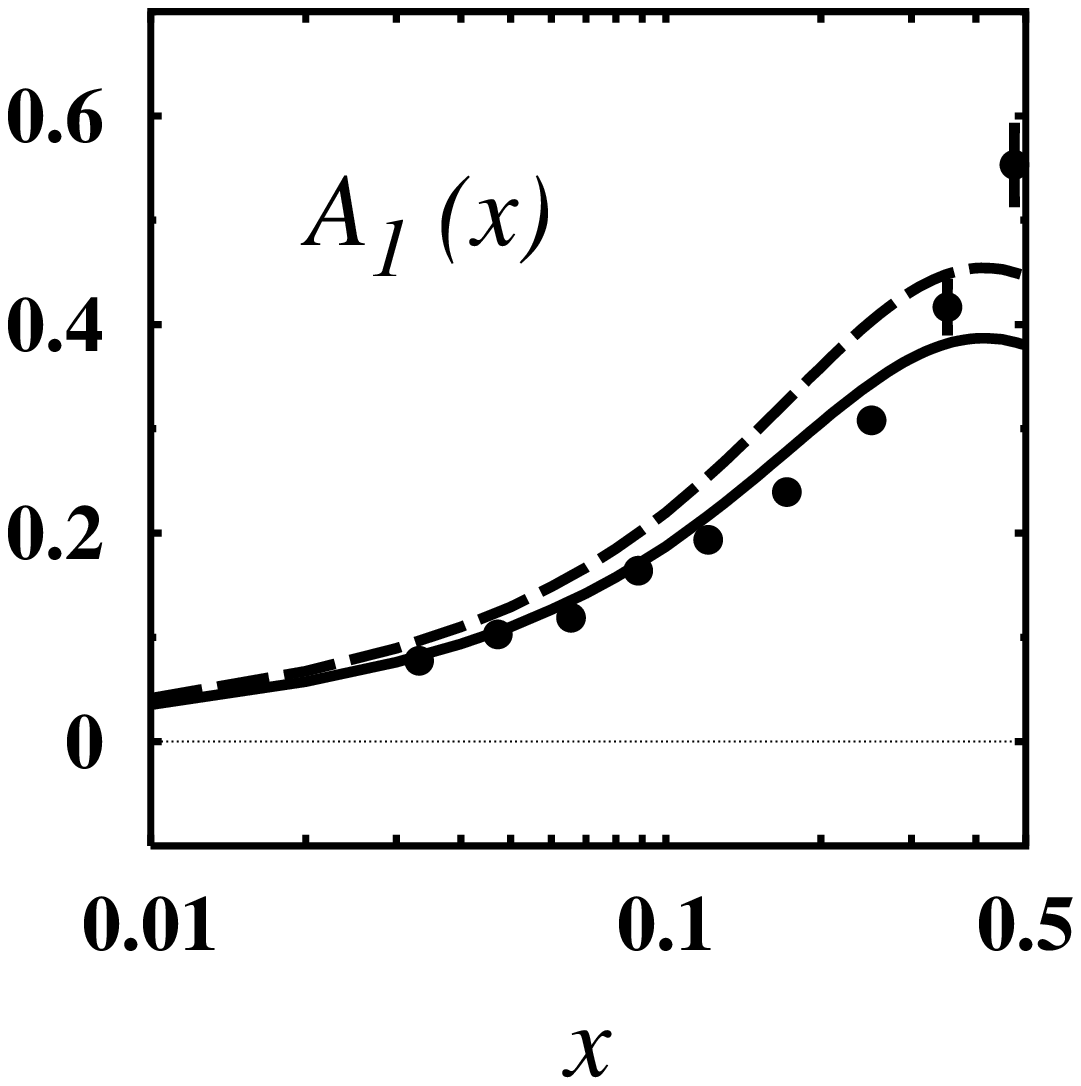}
\end{center}
\caption[]{\it The inclusive spin asymmetry, Eq.(\ref{A1_incl}), 
evaluated with the GRSV/GRV parametrization for the 
polarized/unpolarized parton distributions, at $Q^2 = 2.5\, {\rm GeV}^2$.
\underline{Solid line:} Eq.(\ref{A1_incl}) evaluated without 
the explicit factor $[1 + R(x, Q^2)]$. \underline{Dashed line:}
same as solid line, but explicitly including the factor
$[1 + R(x, Q^2)]$. {\it Dots:} HERMES data from Ref.\cite{HERMES99}.}
\label{fig_incl}
\end{figure}
\par
The decomposition Eq.(\ref{A_decomposition}) now allows us to consistently
combine the information available from inclusive DIS,
contained in the standard parametrizations of polarized parton 
distributions, with the results of our model calculation of the
flavor asymmetries of the antiquark distributions, when computing
the total semi-inclusive spin asymmetry. To evaluate the numerators of the
contributions $A_{1, \, u}^h, A_{1, \, d}^h$, and $A_{1, \, s}^h$ we use 
the GRSV LO parametrizations of 
the distributions $\Delta u(x) + \Delta \bar u(x), \;
\Delta d(x) + \Delta \bar d(x)$, $\Delta s(x) + \Delta \bar s(x)$ 
\cite{GRSV96}.\footnote{To be consistent
with our treatment of $SU(3)$ flavor symmetry breaking in the model
calculation we take the so-called ``standard'' scenario.}
For the contribution $A_{1, \, 0}^h$ involving the flavor--singlet 
antiquark distribution, 
$\Delta \bar u(x) + \Delta \bar d(x) + \Delta \bar s(x)$,
we also use the GRSV LO parametrization, which is in
good agreement with the result of the calculation in the 
chiral quark--soliton model \cite{WK99}.
To estimate the contributions of the flavor--asymmetric antiquark 
distributions to the spin asymmetry, 
$A_{1, \, 3}^h$ and $A_{1, \, 8}^h$, we take the
results of the calculation in the chiral quark--soliton model, 
{\it cf.}\ Fig.\ref{fig_delta} and Eq.(\ref{8_from_3}), evolved 
from the scale of $\mu^2 = (600\, {\rm MeV} )^2$ up to 
the experimental scale. Note that in leading order
$\Delta \bar u(x) - \Delta \bar d(x)$  and
$\Delta \bar u(x) + \Delta \bar d(x) - 2 \Delta \bar s(x)$ 
do not mix with the total distributions under evolution
[we neglect $SU(3)$--symmetry breaking effects due to the 
finite strange quark mass in the evolution].
Finally, to evaluate the denominators in Eqs.(\ref{X_u})--(\ref{X_8}), 
we use the GRV LO parametrization of unpolarized parton 
distributions \cite{GRV95},
which agrees well with the model calculations of 
Refs.\cite{DPPPW96,DPPPW97,PPGWW98,WK99}.
\par
We emphasize that with the above ``hybrid'' set of polarized 
parton distributions we automatically fit {\it all inclusive data}, 
since the GRSV
parametrization was determined from fits to the structure function
data. In particular, the GRSV parametrization describes well the 
HERMES data for the inclusive spin asymmetry, Eq.(\ref{A1_incl}), as 
can be seen from Fig.\ref{fig_incl}.\footnote{The 
GRSV parametrization was derived by fitting to the ratio Eq.(\ref{A1_incl})
including the factor $[1 + R(x, Q^2)]$, so that this factor is contained
in the polarized parton distribution functions and should not be
included explicitly when evaluating Eq.(\ref{A1_incl}). The effect
of this factor is shown in Fig.\ref{fig_incl}.} Note that
the HERMES data are in good agreement with the SLAC E143 data
for the inclusive spin asymmetry \cite{E143}.
\par
{\it Spin asymmetries in charged pion production.}\
In order to determine the contributions of the various combinations of 
polarized quark and antiquark distributions to the semi-inclusive spin 
asymmetry, Eq.(\ref{A1_x}), we need to evaluate the coefficients
$X_u^h, \ldots X_8^h$, Eqs.(\ref{X_u})--(\ref{X_8}), using a set of
quark and antiquark 
fragmentation functions. Like the parton distributions, the fragmentation 
functions are process--independent quantities, which can in principle
be determined from a variety of hard processes with hadronic final states,
such as $e^+e^-$ annihilation, or in semi-inclusive hadron production in 
unpolarized DIS. Unfortunately, experimental knowledge of fragmentation 
functions is still comparatively poor. This applies in particular to the
so-called unfavored fragmentation functions, which we need to study the 
influence of flavor symmetry in the antiquark distribution, see
Eqs.(\ref{X_3}) and (\ref{X_8}). The low--scale parametrizations of 
unpolarized 
quark and antiquark fragmentation functions of Binnewies {\it et al}., 
which fit variety of $e^+e^-$ annihilation and semi-inclusive DIS data,
describe only fragmentation into positively and negatively charged 
particles combined, which is not sufficient for our purposes. 
The $\pi^+$ and $\pi^-$ fragmentation functions separately have
been extracted at low $Q^2$ from semi-inclusive pion production at 
HERMES \cite{Geiger98}. 
The number of independent pion fragmentation functions 
is reduced by isospin and charge conjugation invariance:
\be
D_u^{\pi+} \;\; =\;\;  D_{\bar d}^{\pi+} \;\; = \;\; D_d^{\pi-}
\;\; = \;\; D_{\bar u}^{\pi-} &\equiv & D ,
\\[1ex]
D_d^{\pi+} \;\; =\;\;  D_{\bar u}^{\pi+} \;\; = \;\; D_u^{\pi-}
\;\; = \;\; D_{\bar d}^{\pi-} &\equiv & \widetilde{D} ,
\ee
where $D$ and $\widetilde{D}$ are called, respectively, favored and
unfavored fragmentation function. In addition, in Ref.\cite{Geiger98}
it was assumed that the strange quark fragmentation
function into pions is approximately equal to the unfavored 
fragmentation function for $u$ and $d$ quarks:
\be
D_s^{\pi+} \;\; =\;\;  D_{\bar s}^{\pi-} \;\; 
\approx \;\; D_{\bar s}^{\pi+} \;\; =\;\;  D_s^{\pi-} \;\; 
&\approx& \widetilde{D} .
\ee
With these assumptions we can compute the integrals 
Eq.(\ref{D}) for $h = \pi^+, \pi^-$ using the HERMES fragmentation
functions
(extraction method 1, corrected for $4 \pi$ acceptance) \cite{Geiger98}. 
The values of the coefficients $X_u^{\pi\pm}, \ldots 
X_8^{\pi\pm}$ integrals obtained with a cutoff 
$z_{\rm min} = 0.2$ (the value used 
in the analysis of HERMES charged hadron data \cite{HERMES99}) 
are given in rows 1 and 2 of Table \ref{tab_X}.
%
%
\begin{table}
\begin{center}
\begin{tabular}{|r|r|r|r|r|r|r|} \hline
      & $X_u^h$ & $X_d^h$ & $X_s^h$ & $X_0^h$ & $X_3^h$ & $X_8^h$ \\ 
\hline
$h = \pi^+$ & 0.200 & 0.029 & 0.029 & -0.021 & -0.053 & -0.011 \\
    $\pi^-$ & 0.115 & 0.050 & 0.029 &  0.021 &   0.053 & 0.011 \\
\hline
$h = h^+$   & 0.277 & 0.040 & 0.035 & -0.037 & -0.077 & -0.023 \\
    $h^-$   & 0.140 & 0.056 & 0.043 & 0.037 & 0.077 & 0.023 \\
\hline
$h = K^+$   & 0.049 & 0.004 & 0.004 & -0.008 & -0.017 & -0.008 \\
    $K^-$   & 0.016 & 0.004 & 0.012 &  0.008 & 0.017 & 0.008 \\
\hline
\end{tabular}
\end{center}
\caption{\it \underline{Rows 1 and 2:} 
The coefficients Eq.(\ref{X_u})--(\ref{X_8}) 
for $\pi^+$ and $\pi^-$ production, evaluated with the HERMES 
fragmentation functions \cite{Geiger98} 
for $z_{\rm min} = 0.2$. \underline{Rows 3 and 4:} 
Same for charged hadron production. The kaon and proton 
fragmentation functions have been 
taken from the EMC measurements \cite{EMC89} (details see text).
\underline{Rows 5 and 6:} Same for kaon production only.}
\label{tab_X}
\end{table}
The numerical values of the coefficients
reveal two things: First, the dominant contribution 
to the charged hadron asymmetry comes from the sum of the 
polarized $u$--quark and antiquark distributions in the target, 
$\Delta u(x) + \Delta \bar u (x)$, which is a consequence of the 
large squared charge of the $u$--quark. Second, among the various flavor
combinations of the antiquark distributions the isovector one,
$\Delta \bar u(x) - \Delta \bar d (x)$ enters with the largest
coefficient --- a fortunate circumstance for attempts to extract
this distribution from the data.
\par
\begin{figure}[t]
\begin{tabular}{rr}
\includegraphics[width=7.5cm,height=7.5cm]{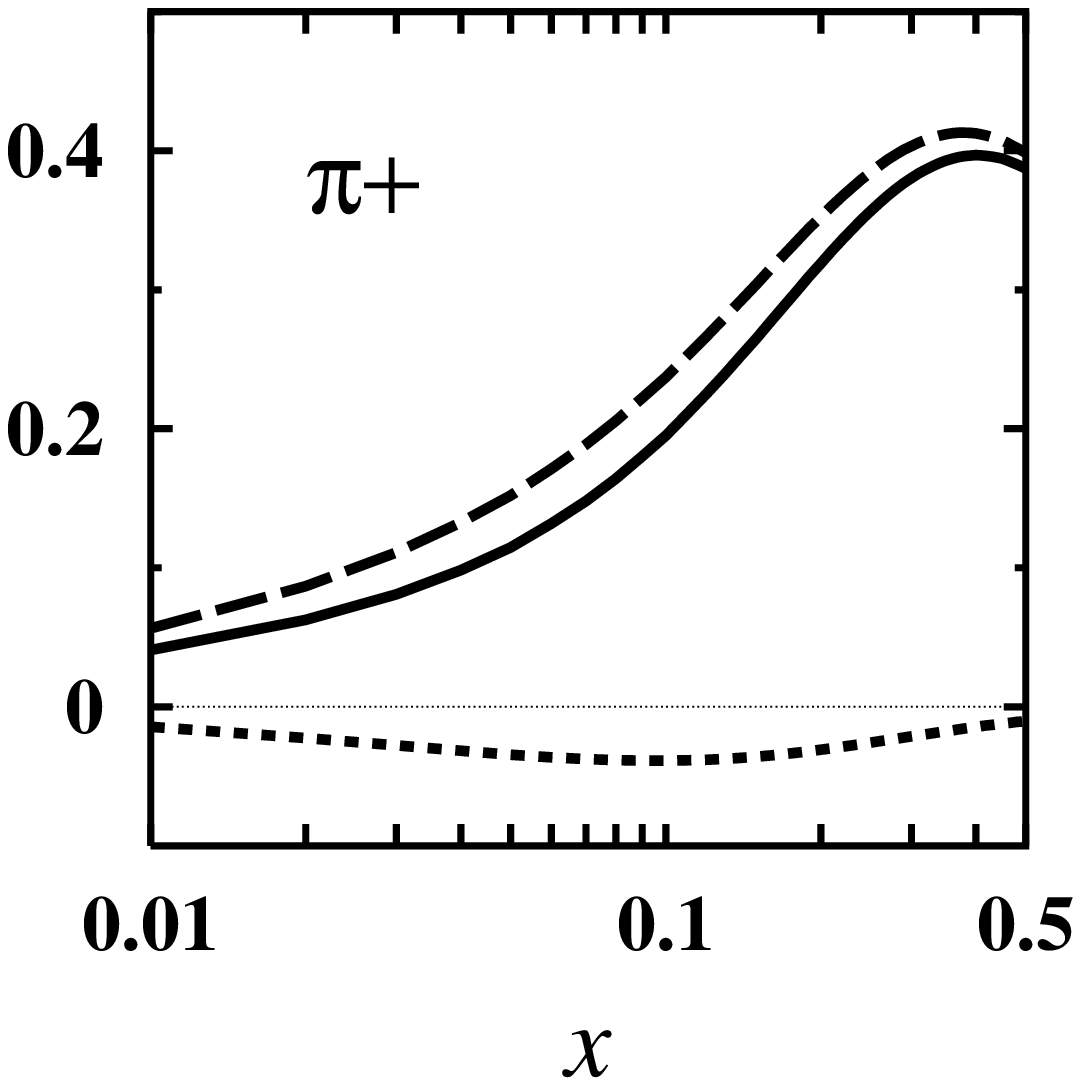}
&
\includegraphics[width=7.5cm,height=7.5cm]{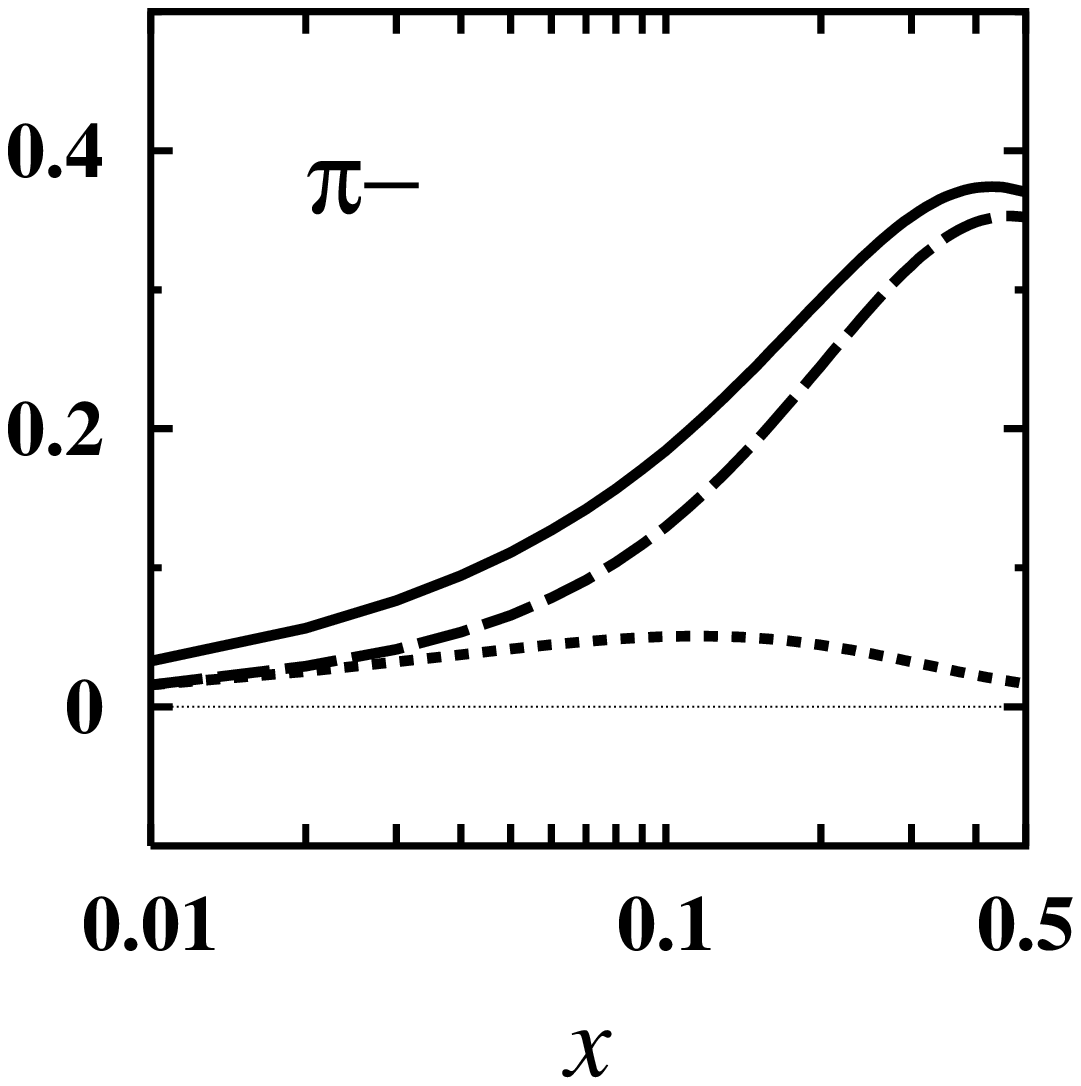}
\end{tabular}
\caption[]{\it The spin asymmetries for $\pi^+$ and $\pi^-$ production in 
semi-inclusive DIS off the proton ($Q^2 = 2.5\, {\rm GeV}^2$, 
$z_{\rm min} = 0.2$). \underline{Dashed lines:} Sum of contributions 
$A_{1, \, u}^{\pi+}$, $A_{1, \, d}^{\pi+}, A_{1, \, s}^{\pi+}$, and 
$A_{1, \, 0}^{\pi+}$
(and respectively for $\pi-$), {\it cf.}\ Eq.(\ref{A_decomposition}). 
\underline{Dotted lines:} 
Contributions $A_{1, \, 3}^{\pi+}$ and $A_{1, \, 3}^{\pi -}$, 
respectively, proportional to the flavor asymmetry
$\Delta \bar u (x) - \Delta \bar d (x)$ in the target, 
evaluated with the distribution shown in Fig.\ref{fig_delta}.
\underline{Solid lines:} Total results, including the effect
of $\Delta \bar u (x) - \Delta \bar d (x)$. The contributions
$A_{1, \, 8}^{\pi +}$ and $A_{1, \, 8}^{\pi -}$ are very small
and therefore not shown separately.}
\label{fig_pipm}
\end{figure}

The results for the spin asymmetry in semi-inclusive $\pi^+$ and $\pi^-$
production are shown in Fig.\ref{fig_pipm}.
The dashed lines show the results obtained taking into account 
only the contributions $A_{1, \, u}^{\pi\pm}, A_{1, \, d}^{\pi\pm}, 
A_{1, \, s}^{\pi\pm}$ and $A_{1, \, 0}^{\pi\pm}$ to the spin asymmetry,
Eq.(\ref{A_decomposition}), {\it i.e.}, what 
would be obtained without flavor asymmetry in the polarized
antiquark distribution of the proton. The contributions 
$A_{1, \, 3}^{\pi\pm}$,
proportional to $\Delta \bar u (x) - \Delta \bar d (x)$, 
are shown by the dotted lines. The total results for the 
asymmetries, including all contributions, are shown by the 
the solid lines. [The contributions $A_{1, \, 8}^{\pi\pm}$ are very small, 
of the order of 10\% of $A_{1, \, 3}^{\pi\pm}$, and not shown separately.]
One sees that in both cases the effect of the flavor asymmetry
of the antiquark distribution is noticable. 
%
%
\par
{\it Spin asymmetries in charged hadron production and comparison with 
the HERMES data.}\ We now
turn to the spin asymmetries in charged production, which have
been measured by the SMC \cite{SMC98} and HERMES \cite{HERMES99}
experiments. Unfortunately, no complete set of quark 
fragmentation functions for charged hadrons ($K^+, K^-, p, \bar p$)
at the HERMES scale ($Q^2 = 2.5\, {\rm GeV}^2$) is available. We 
therefore take recourse to the older EMC results for the fragmentation 
functions \cite{EMC89}, in which positively and negatively charged 
hadrons were separated. 
These data have been taken at a higher scale of
$Q^2 = 25\, {\rm GeV}^2$. Since it turns out that the dominant 
contribution to the semi-inclusive spin asymmetry for production
of charged hadrons comes from the pions,
we combine the HERMES result for the pion fragmentation functions
\cite{Geiger98} with the EMC fragmentation functions for kaons and 
protons \cite{EMC89}, ignoring the scale dependence of 
the kaon and proton fragmentation functions, which anyway give a small 
contribution.\footnote{In 
principle the evolution equations for 
fragmentation functions \cite{GL72,Stratmann97} would allow us 
to parametrize the EMC results
in terms of fragmentation functions at a lower scale; however, in
order to do so consistently we would need also the gluon fragmentation
functions at the higher scale, which has not been measured by EMC.}
Again, isospin and charge conjugation allow us to write:
\be
D_u^{K+} \;\; =\;\;  D_{\bar u}^{K-} &\equiv & D^K ,
\\[1ex]
D_d^{K+} \;\; =\;\;  D_{\bar d}^{K-} &\equiv & \widetilde{D}^K ,
\\[1ex]
D_u^{p} \;\; =\;\;  D_{\bar u}^{\bar p} &\equiv & D^p ,
\\[1ex]
D_u^{\bar p} \;\; =\;\;  D_{\bar u}^{p} &\equiv & \widetilde{D}^p .
\ee
Furthermore, following the analysis in Ref.\cite{EMC89}, we shall assume 
that
\be
D_s^{K-} \;\; =\;\; D_{\bar s}^{K+} &\approx & D^K ,
\\[1ex]
D_s^{K+} \;\; =\;\;  D_{\bar s}^{K-} &\approx & \widetilde{D}^K ,
\\[1ex]
D_d^p \;\; = \;\;  D_{\bar d}^{\bar p} &\approx & D^p ,
\\[1ex]
D_s^p \;\; = \;\;  D_{\bar s}^{\bar p} 
\;\; \approx \;\; D_{\bar d}^p \;\; = \;\;  D_d^{\bar p} 
&=& \widetilde{D}^p .
\ee
With these assumptions all relevant fragmentation functions for 
$h+ \approx \pi^+ + K^+ + p$ and $h- \approx  \pi^- + K^- + \bar p$ 
can be estimated in terms of the six functions 
$D, \widetilde{D}, D^K, \widetilde{D}^K, D^p$ and $\widetilde{D}^p$.
Evaluating the integrals with $z_{\rm min} = 0.2$ (the cutoff used in 
the analysis of HERMES data \cite{HERMES99}) we obtain the values
shown in rows 3 and 4 of Table \ref{tab_X}. One sees that the
vaules are not too different from those obtained for $\pi^+$ and 
$\pi^-$ production;
only the sensitivity to the strange quark distributions has increased
somewhat due to the inclusion of kaon production.
%
%
\begin{figure}[t]
\begin{tabular}{rr}
\includegraphics[width=7.5cm,height=7.5cm]{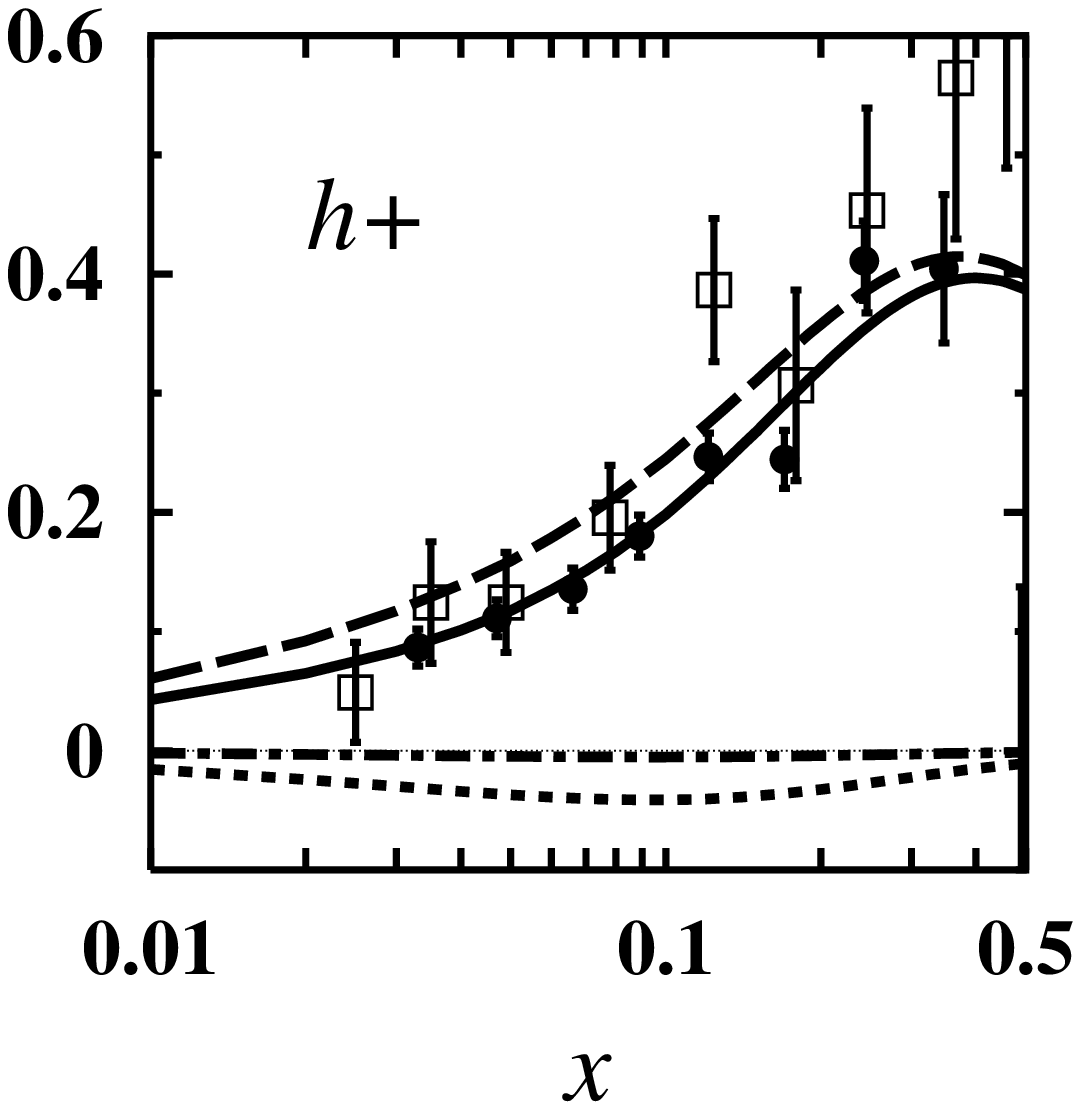}
&
\includegraphics[width=7.5cm,height=7.5cm]{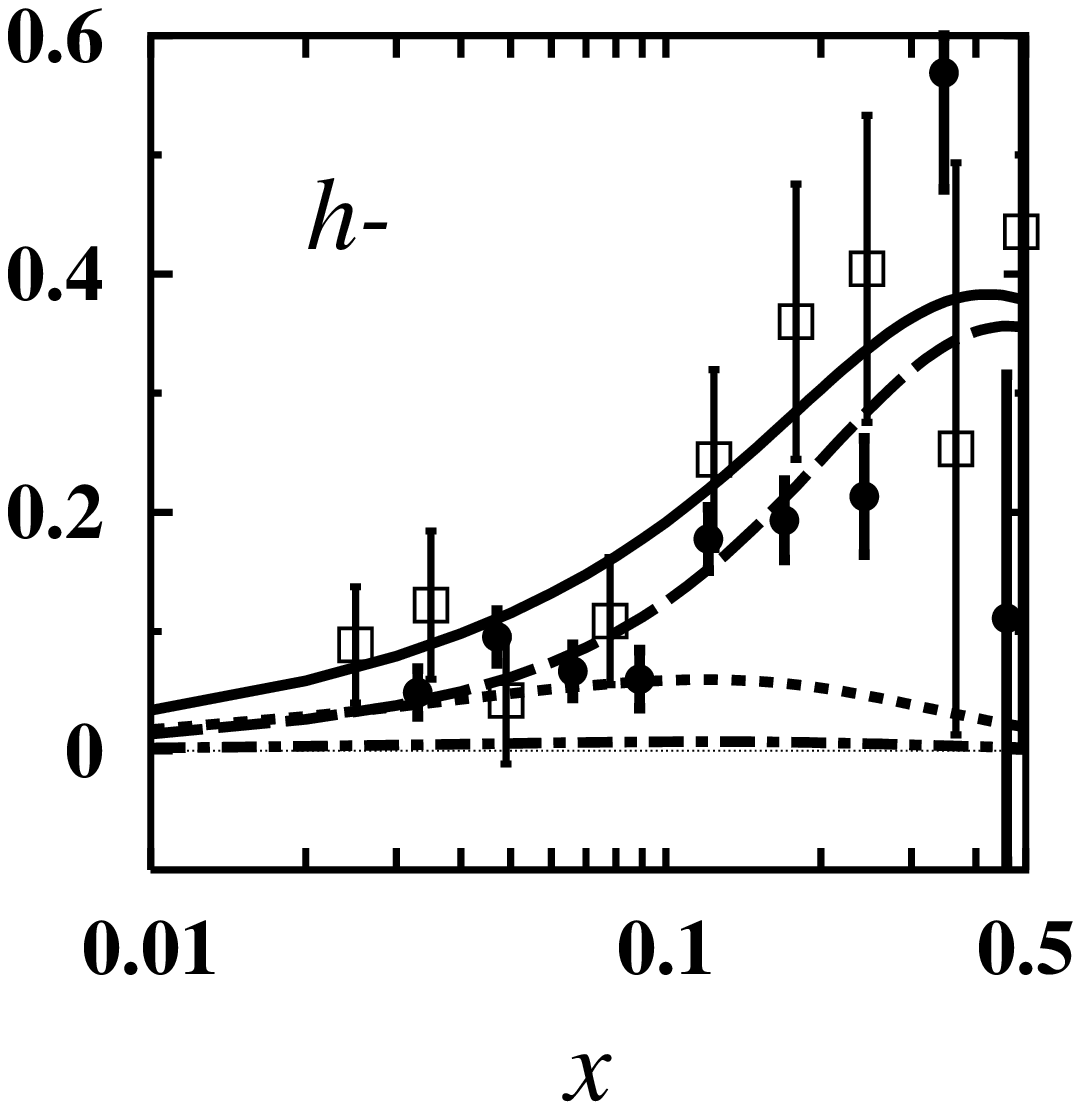}
\end{tabular}
\caption[]{\it The spin asymmetries for $h^+$ and $h^-$ production in 
semi-inclusive DIS off the proton ($Q^2 = 2.5\, {\rm GeV}^2$, 
$z_{\rm min} = 0.2$). \underline{Dashed lines:} Sum of contributions 
$A_{1, \, u}^{h+}$, $A_{1, \, d}^{h+}, A_{1, \, s}^{h+}$, and 
$A_{1, \, 0}^{h+}$
(and respectively for $h^-$), {\it cf.}\ Eq.(\ref{A_decomposition}). 
\underline{Dotted lines:} Contributions $A_{1, \, 3}^{h+}$ and 
$A_{1, \, 3}^{h-}$, respectively, proportional to the flavor asymmetry
$\Delta \bar u (x) - \Delta \bar d (x)$ in the target, 
evaluated with the distribution shown in Fig.\ref{fig_delta}.
\underline{Dash--dotted line:} Contributions $A_{1, \, 8}^{h+}$ 
and $A_{1, \, 8}^{h-}$, respectively. 
\underline{Solid lines:} Total results, including the effect
of flavor asymmetry of the polarized antiquark distribution. 
\underline{Open Squares:} SMC data \cite{SMC98}.
\underline{Filled Circles:} HERMES data \cite{HERMES99}.}
\label{fig_hpm}
\end{figure}
\par
In Fig.\ref{fig_hpm} we show the results for 
the spin asymmetries in $h^+$ and $h^-$ production. As in 
Fig.\ref{fig_pipm} for $\pi^+$ and $\pi-$ we plot
the asymmetry that would be obtained without flavor asymmetry of the
polarized antiquark distribution in the target 
($A_{1, \, u}^{h\pm} + A_{1, \, d}^{h\pm} 
+ A_{1, \, s}^{h\pm} + A_{1, \, 0}^{h\pm}$), the contributions 
$A_{1, \, 3}^{h\pm}$ and $A_{1, \, 8}^{h\pm}$ containing the
effect of the flavor asymmetry, and the total result.
\par
A preliminary comparison of the theoretical results for $A_1^{h\pm}$
shows that the spin asymmetries computed including the effects 
of $\Delta \bar u (x) - \Delta \bar d (x)$ (solid lines in 
Fig.\ref{fig_hpm}) are consistent with the HERMES \cite{HERMES99} 
and SMC \cite{SMC98} data. The accuracy of the present data, 
in particular for $A_1^{h-}$, seems not to be sufficient for a 
definite choice between the theoretical results obtained with 
(solid lines) and without (dashed lines in Fig.\ref{fig_hpm}) 
the flavor asymmetry of the polarized antiquark distribution. 
Also, one should be aware that there are several sources of uncertainty
in our theoretical predictions. The greatest uncertainty comes from
our imperfect knowledge of the fragmentation function, mostly from
the pion fragmentation functions. 
In the analysis of Ref.\cite{Geiger98}, significant 
corrections were applied to the measured fragmentation functions
in order to compensate
for the acceptance of the HERMES detector. (We use the $4\pi$--corrected
fragmentation functions in our above estimates.) A full error analysis 
would require keeping
track of the systematic error in the fragmentation functions
and is outside the scope of this
paper. One should consider the possibility of changing the above analysis
such as to be able to work directly with the fragmentation functions
specific to the HERMES 
detector, avoiding the $4\pi$ corrections. It is conceivable that in this 
way one could significantly reduce the systematic error in 
the fragmentation functions.
\par
Recently, Morii and Yamanishi have attempted to extract 
$\Delta\bar u (x) - \Delta\bar d (x)$ from the data for
polarized semi-inclusive asymmetries by combining 
data taken with proton and Helium targets \cite{Morii99}. 
(It was shown in Ref.\cite{WW99} that the asymmetry calculated
in the chiral quark--soliton model is consistent with 
their bounds obtained in Ref.\cite{Morii99}.)
Since in the case of the HERMES experiment the statistics
of the Helium data is significantly worse than for the 
proton, this combination of data results in a loss of accuracy. 
Also, this 
approach requires accurate compensation for nuclear binding effects.
In contrast, our method of analysis relies on proton data 
only. 
%
%
\begin{figure}[t]
\begin{tabular}{rr}
\includegraphics[width=7.5cm,height=7.5cm]{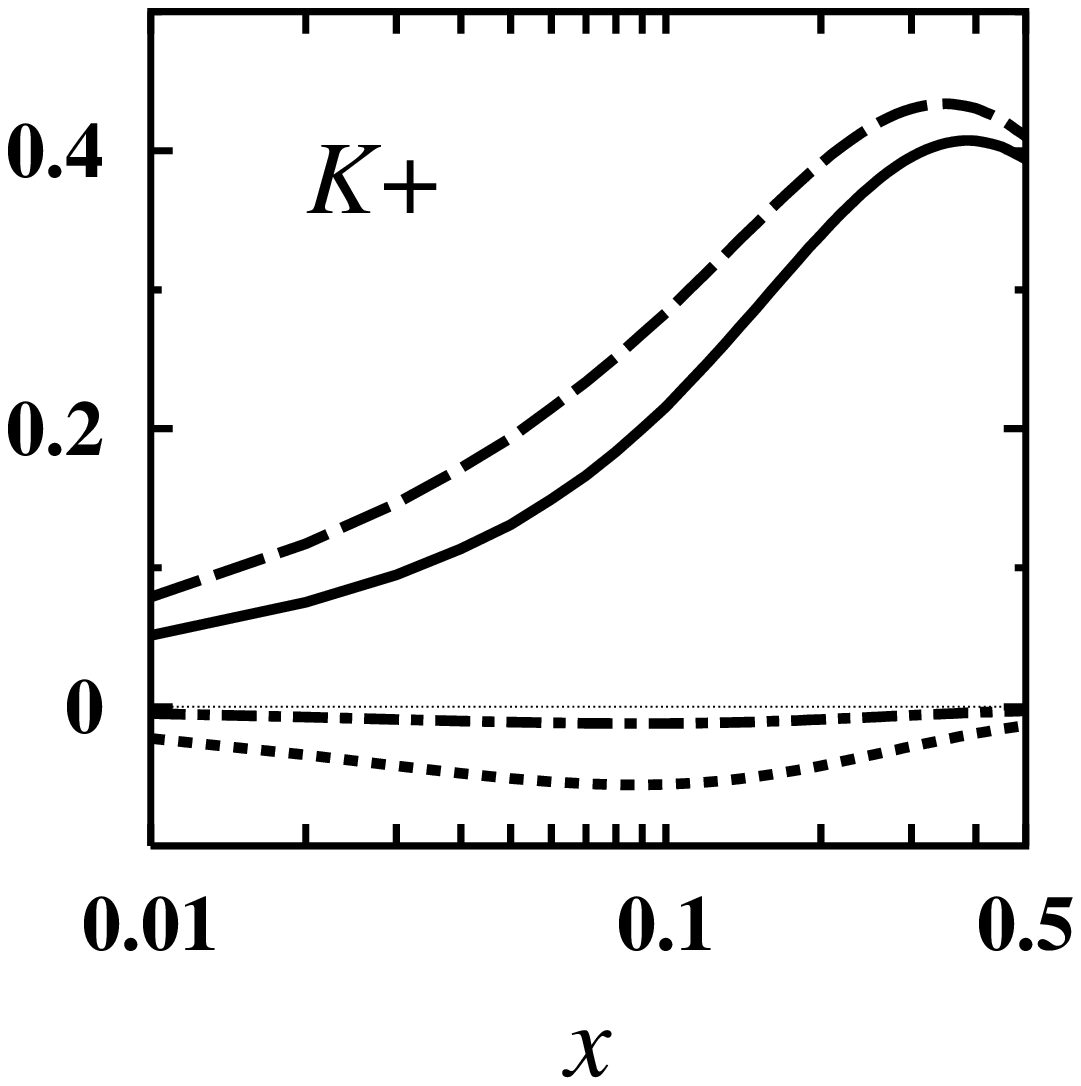}
&
\includegraphics[width=7.5cm,height=7.5cm]{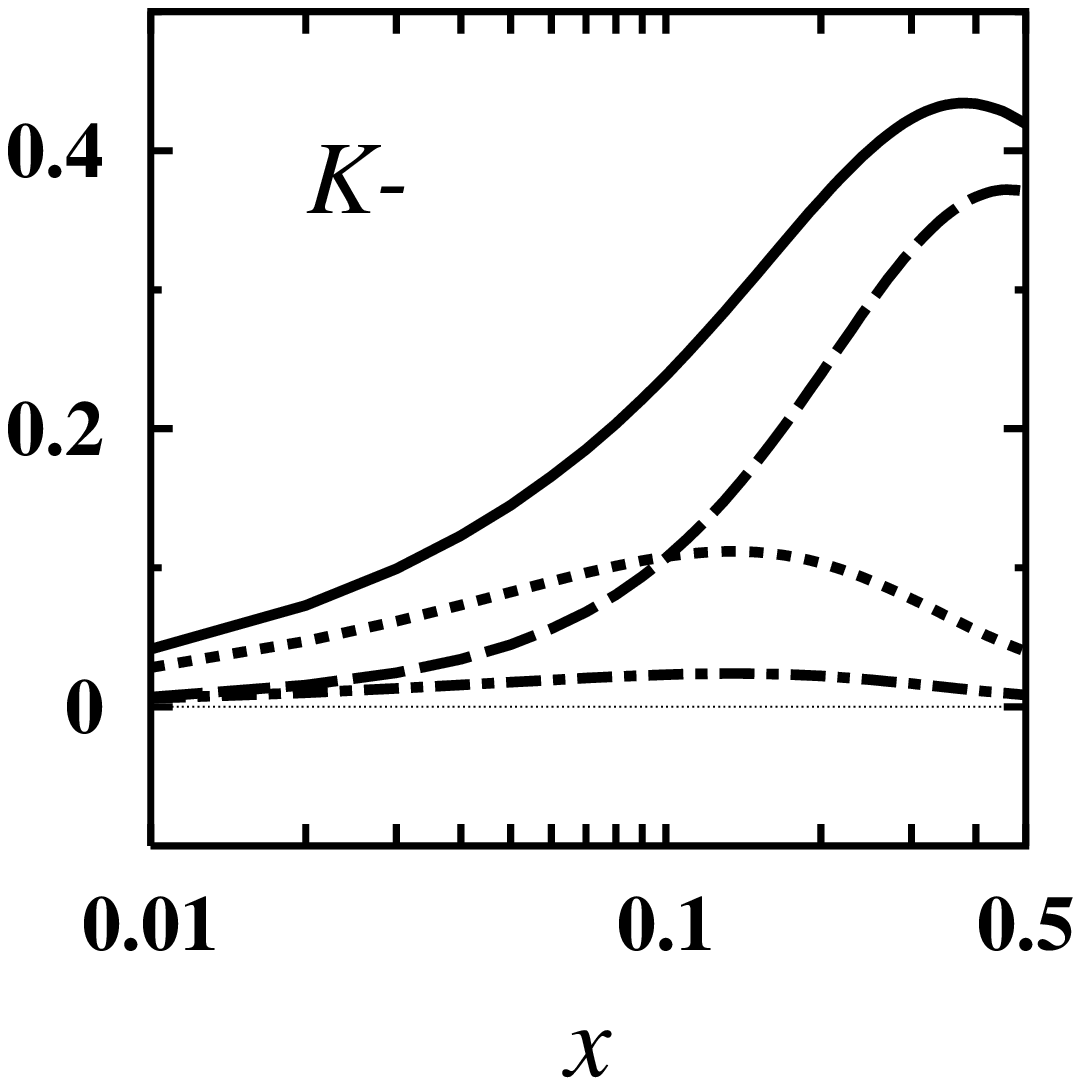}
\end{tabular}
\caption[]{\it The spin asymmetries for $K^+$ and $K^-$ production in 
semi-inclusive DIS off the proton ($Q^2 = 2.5\, {\rm GeV}^2$, 
$z_{\rm min} = 0.2$). {\it Dashed lines:} Sum of contributions 
$A_{1, \, u}^{K+}$, $A_{1, \, d}^{K+}, A_{1, \, s}^{K+}$, and 
$A_{1, \, 0}^{K+}$
(and respectively for $K^-$), {\it cf.}\ Eq.(\ref{A_decomposition}). 
\underline{Dotted lines:} Contributions $A_{1, \, 3}^{K+}$ and 
$A_{1, \, 3}^{K-}$, respectively, proportional to the flavor asymmetry
$\Delta \bar u (x) - \Delta \bar d (x)$ in the target, 
evaluated with the distribution shown in Fig.\ref{fig_delta}.
Note the large contribution in $K^-$ production.
\underline{Dash--dotted line:} Contributions $A_{1, \, 8}^{K+}$ 
and $A_{1, \, 8}^{K-}$, respectively. 
\underline{Solid lines:} Total results, including the effect
of flavor asymmetry of the polarized antiquark distribution.}
\label{fig_kpm}
\end{figure}
\par
{\it Spin asymmetries in charged kaon production.} It is interesting 
to consider separately also the spin asymmetries in the production 
of charged kaons 
only. In particular, $K^-$ cannot be produced by favored 
fragmentation of either $u$ or $d$ quarks in the target, which 
makes for the bulk contribution to the semi-inclusive spin asymmetry
in $\pi^\pm$ or $h^\pm$ production. In this case one might expect 
a large sensitivity of the spin asymmetry to the flavor asymmetries 
of the polarized antiquark distributions. For a rough estimate we can 
use the EMC fragmentation functions
to evaluate the coefficients Eq.(\ref{X_u})--(\ref{X_8}) for 
$K^+$ and $K^-$; the results are shown in rows 5 and 6 of
Table \ref{tab_X}.
The contributions to the spin asymmetries are shown in 
Fig.\ref{fig_kpm}. One sees that, in particular
in the case of $K^-$ production, the contribution proportional
to $\Delta \bar u (x) - \Delta \bar d (x)$ in the proton
is large. Thus, semi-inclusive charged kaon production 
could be a sensitive test of the flavor decomposition of the antiquark
distribution in the nucleon.
\section{Conclusions and Outlook} 
Starting from the observation that the $1/N_c$--expansion 
predicts large flavor asymmetries of the polarized antiquark 
distributions, and the quantitative estimates for
$\Delta\bar u(x) - \Delta\bar d (x)$ and 
$\Delta\bar u(x) + \Delta\bar d (x) - 2 \Delta \bar s (x)$
obtained from the chiral quark--soliton model, we have explored several
consequences of a large flavor asymmetry of the polarized antiquark
distributions.
\par
On the theoretical side, we have argued that the very small value for 
the polarized antiquark flavor asymmetry obtained from polarized rho 
meson exchange in the meson cloud picture \cite{Fries98} does not rule 
out a large asymmetry, since polarized 
rho meson exchange is by far not the dominant contribution 
to $\Delta \bar u(x) - \Delta \bar d (x)$ in that
approach. Comparison with the large--$N_c$ result suggests
that, in the terms of the meson cloud model, a large contribution 
is likely to come from the interference of pion and ``sigma meson'' 
exchange. It could be interesting to explore this 
qualitative suggestion in more detail within the meson
cloud picture.
\par
As to experimental consequences, we have found that the large flavor 
asymmetry predicted by the chiral quark soliton model is
consistent with the recent HERMES data on
spin asymmetries in semi-inclusive charged hadron production.
Our conclusions are based on the presently available information
on quark and antiquark fragmentation functions; however, to the extent 
that we have explored it, they seem to be robust with regard to
the systematic uncertainties in the fragmentation functions.
We do not claim that at the present level of accuracy the
HERMES data for $A^{h\pm}_1$ necessitate a large flavor asymmetry
of the antiquark distribution in the proton. However, assuming
a large flavor asymmetry we obtain a good fit to the data.
In this sense the HERMES results should not be seen as
evidence for a small flavor asymmetry of the polarized antiquark 
distribution. Also, we have argued that the assumption 
$\Delta \bar u (x) / \bar u (x) = \Delta \bar d (x) / \bar d (x)$
made in the analysis of the HERMES data in Ref.\cite{HERMES99}
artificially limits the contributions from the polarized antiquark 
flavor asymmetry, and thus does not constitute a real alternative
to the reference fit assuming zero flavor asymmetry.
\par
At present, one source of uncertainty in our theoretical results
are the systematic errors in the HERMES quark and antiquark fragmentation 
functions introduced by the corrections for $4\pi$ acceptance. It
would be worthwhile to investigate if not the above analysis
could be carried out directly with the fragmentation functions
for HERMES acceptance, which could considerably reduce the systematic 
error.
\par
We have shown that the flavor asymmetry of the polarized antiquark 
distribution makes a particularly large contribution to the
spin asymmetries in charged kaon production. Such measurements
could be an interesting option with detectors
which allow discrimination between pions and kaons in the 
final state, {\it e.g.}\ at HERMES or CEBAF.
\\[.5cm]
{\large\bf Acknowledgements} \\[.3cm]
The authors are grateful to A. Sch\"afer and M. Strikman, from
discussions with whom arose the idea to write this note,
and to M.\ D\"uren for comments. Helpful hints from Ph.\ Geiger,
R. Jakob, B. Kniehl, P.V.\ Pobylitsa, and P. Schweitzer are 
also gratefully acknowledged.
\\
This work has been supported in part by the Deutsche 
Forschungsgemeinschaft (DFG), by a joint grant of the 
DFG and the Russian Foundation for Basic Research, by
the German Ministry of Education and Research (BMBF),
and by COSY, J\"ulich.
\end{document}